\documentclass[12pt]{iopart}
\usepackage{bm,color}

\makeatletter
\def\oversortoftilde#1{\mathop{\vbox{\m@th\ialign{##\crcr\noalign{\kern3\p@}%
      \sortoftildefill\crcr\noalign{\kern3\p@\nointerlineskip}%
      $\hfil\displaystyle{#1}\hfil$\crcr}}}\limits}

\def\sortoftildefill{$\m@th \setbox\z@\hbox{$\braceld$}%
  \braceld\leaders\vrule \@height\ht\z@ \@depth\z@\hfill\braceru$}

\makeatother

\begin{document}
\title{Hierarchy of second-order gyrokinetic Hamiltonian models for particle-in-cell codes}

\author{N Tronko $^{1,2}$, A Bottino $^1$, C Chandre $^3$ and E Sonnendruecker $^{1,2}$}

\address{$^1$ Max-Planck Institute for Plasma Physics, 85748, Garching, Germany,
$^2$TU Munich, Mathematics Center, 85747, Garching, Germany, $^3$ Aix Marseille Univ, CNRS, Centrale Marseille, I2M, Marseille, France}

\begin{abstract}
The reduced-particle model is the central element for the systematic derivation of the gyrokinetic Vlasov-Maxwell equations from first principles. Coupled to the fields inside the gyrokinetic field-particle Lagrangian, the reduced-particle model defines polarization and magnetization effects appearing in the gyrokinetic Maxwell equations. It is also used for the reconstruction of the gyrokinetic Vlasov equation from the particle characteristics.
Various representations of reduced-particle models are available according to the choice of the gyrokinetic phase space coordinates. In this paper, the Hamiltonian representation of the reduced particle dynamics at an order suitable for the implementation in particle-in-cell simulations is explicitly derived from the general reduction procedure.
The second-order (with respect to the fluctuating electromagnetic fields), full Finite Larmor Radius (FLR) Hamiltonian gyrokinetic particle model as well as the second-order model suitable specifically for the \textit{long-wavelength} approximation (i.e., containing up to the second-order FLR corrections), are derived and compared to the model recently implemented in the particle-in-cell code ORB5.
We show that the same \textit{long-wavelength} approximate equations can also be derived by taking the proper limit of the full FLR model.
\end{abstract}
\maketitle
\section{Introduction}

A magnetized plasma represents a complex system with multi-scale dynamics in both time and space, which is a challenge for numerical implementation.
For several decades the gyrokinetic dynamical reduction \cite{Frieman_Chen_1982}, \cite{Littlejohn_1983}, \cite{Brizard_Hahm} has been of interest as one of the most powerful tools to study this multi-scale problem both numerically and analytically (see for example \cite{Garbet_Idomura_2010},\cite{TBG_2017}). 

One of the main challenges for a systematic derivation of the gyrokinetic theory comes from the fact that the general derivation \cite{Brizard_Hahm} includes several groups of small parameters. The first group is connected to the relative amplitude of the background inhomogeneities, while the second group originates from the ratio between the amplitude of fluctuating electromagnetic fields and background quantities. The full gyrokinetic derivation considers both kinds of small parameters of the same order, which makes not only the derivation of the final model but also its numerical implementation rather challenging.

Nowadays, a simplified ordering, which allows one to transfer corrections from the background inhomogeneities at the next higher order in perturbation 
theory, is widely used for the derivation of models implemented in gyrokinetic codes.
In particular, the Vlasov-Maxwell reduced models  with linearised polarisation and magnetisation are of great interest.

From the perspective of the approximations performed on the electromagnetic field fluctuations, models for global codes (i.e., ORB5 \cite{Jolliet_2007} and GENE \cite{Goerler_2011}) are usually derived in the $\textit{low}-\beta$ approximation, meaning that the parallel component of the perturbed magnetic field is systematically neglected, which is the framework we consider in this article.

The gyrokinetic theory derived from the field theory formulation has a great advantage over the direct asymptotic decomposition of the Vlasov equation first derived in Ref.~\cite{Frieman_Chen_1982}.  In fact, considering the coupling between fields and reduced particle dynamics within the same mathematical framework gives access to the derivation of self-consistently coupled gyrokinetic Vlasov-Maxwell equations. 

The reduced-particle model is an essential element of this derivation. First of all, the reduced particle dynamics affects the Maxwell equations via polarization and magnetization terms. At the same time, it defines the gyrokinetic Vlasov equation, which can be directly reconstructed from its characteristics. The coupling between the reduced field and particle equations can be systematically established via a first-principle derivation from the gyrokinetic Lagrangian \cite{Sugama_2000}, \cite{brizard_prl_2000}. 

A systematic variational framework for gyrokinetic theory has undergone a significant development during the last two decades, while the development of codes started a decade earlier. For this historical reason, some of the major gyrokinetic codes are based on the asymptotic derivation rather than on field theory.
A significant effort toward code verification started in 2014 in the framework of the European project \texttt{VeriGyro}.  In a recent work, \cite{TBS_2016} orderings have been identified and a general gyrokinetic field theory has been developed for an explicit derivation of the model implemented in the PIC code ORB5.

This paper focusses on the detailed derivation of Hamiltonian gyrokinetic models for the reduced particle dynamics suitable for the derivation of the gyrokinetic Vlasov-Maxwell system. Such a derivation has been recently presented in a compact and simplified form for the \textit{long-wavelength} approximation \cite{TBS_2016}.  Here, we extend this derivation by including the full Finite Larmor Radius (FLR) effects in the second-order Hamiltonian model. We show that the very same \textit{long-wavelength} approximated equations can be retrieved by taking the proper limit of the full FLR model.

This paper is organised as follows. In Sec.~\ref{sec:GK_red}, we summarize the main idea of the gyrokinetic reduction. In Sec.~\ref{sec:GK_pert}, we set up the general framework for coordinate change and the derivation of the reduced dynamics. In Sec.~\ref{sec:4}, we present the derivation of the Hamiltonian gyrokinetic models with the full series of FLR corrections and then in Sec.~\ref{sec:5}, in the \textit{long-wavelength} approximation. Finally, in Sec.~\ref{sec:GK_Vlasov}, we give expressions for the corresponding characteristics and provide the corresponding gyrokinetic Vlasov equations. The Vlasov-Maxwell gyrokinetic models, derived from the full FLR second-order gyrokinetic particle model and the model truncated up to the second-order FLR corrections suitable for the \textit{long-wavelength} approximation are currently implemented in the PIC code ORB5 \cite{Jolliet_2007}.

\section{\label{sec:GK_red}Gyrokinetic dynamical reduction in a nutshell}

In magnetised plasmas, the presence of a strong magnetic field induces a separation of the scales of motion. The particle motion is decomposed into a fast rotation around the magnetic field lines and a slow drift motion in the direction perpendicular to the magnetic field. The scale of gyromotion is set by the cyclotron frequency $\Omega=e B/m c$, where $e$ and $m$ are, respectively, the charge and mass of the particles, $B$ is the magnetic field amplitude and $c$ is the speed of light.
The gyromotion is described by a fast gyroangle variable $\theta$ to which corresponds a canonically conjugated, slowly varying magnetic moment $\mu$.
At the lowest order, it is given by 
\begin{equation}
\mu=m v_{\perp}^2/2 B,
\label{eq:mu0}
\end{equation} 
where $v_{\perp}$ is the perpendicular velocity of particle with respect to the magnetic field lines. In slab (constant magnetic field) geometry, $\mu$ is an exact dynamical invariant. However, magnetic curvature effects as well as the presence of electromagnetic fluctuations break this exact invariance.  The gyrokinetic dynamical reduction uses the fact that, averaged over long times, the magnetic moment is conserved, i.e., $\left\langle\dot{\mu}\right\rangle_t=0$.

The goal of the gyrokinetic dynamical reduction consists in building up a new set of phase space variables, such that the $\theta$ dependence is completely uncoupled and $\mu$ has a trivial dynamics, i.e., $\dot{\mu}=0$. Therefore, the reduced particle dynamics  on the $4+1$ dimensional phase space with variables $\left(\mathbf{X},p,\mu\right)$, where $\mathbf{X}$ represents the reduced position,  $p$ is a scalar momentum coordinate and the new magnetic moment $\mu$ is constant. This change of coordinates is constructed via a perturbative series of near-identity phase space transformations, i.e., these transformations are invertible at each step of the perturbative procedure. The reduced position $\bf X$ has a simple geometrical meaning: It represents an instantaneous center of the fast particle rotation around the magnetic field line. Therefore, from the space coordinate point of view the gyrokinetic transformation is a shift between the initial particle coordinate $\bf x$ and the instantaneous center of its rotation $\bf X$. The difference between both positions is the polarization displacement $\bm\rho$, the derivation of which will be discussed later. Performing numerical simulations on the $4+1$ dimensional phase space instead of the $6$ dimensional one and also removal of the fastest time scales enable the drastic reduction of the numerical cost.

The dynamical reduction can be organized in one or two steps. In the framework of the one-step procedure, the contributions from the background geometry non-uniformities and electromagnetic fluctuations to the breaking of the magnetic moment conservation are taken into account simultaneously. The two-step procedure, composed by the {\em guiding-center} reduction and the subsequent  {\em gyrocenter} reduction, allows one to treat those effects separately, which may have some advantages for making a direct link between the coordinate transformation and the polarization effects induced on the reduced particle and field dynamics. Here, we consider the two-step procedure in order to make a clear separation between the polarization contributions associated with each of these transformations at the lowest order.  

In the two-step reduction procedure, a small parameter is associated with each transformation: These parameters are for the {\em guiding-center} $\epsilon_B=\rho_{th}/L_B$, where $\rho_{th}$ is the thermal Larmor radius of particle and $L_B=\left|\bm\nabla B/B\right|^{-1}$ sets the spatial scale for background magnetic field variation, and for the {\em gyrocenter},  $\epsilon_{\delta}=\left(k_{\perp}\rho_{th}\right)\  e\phi_1/T\equiv\epsilon_{\perp}e\phi_1/T$. The dimensionless parameter $\epsilon_{\perp}$ allows one to distinguish between the gyrokinetic theory with $\epsilon_{\perp}\sim \mathcal{O}(1)$ and the drift-kinetic theory with $\epsilon_{\perp}\ll 1$, known also as the \textit{ long-wavelength} approximation when only the second-order $\mathcal{O}(\epsilon_{\perp}^2)$ corrections are included. Following the gyrokinetic ordering relevant for numerical implementation, we consider $\epsilon_B\ll\epsilon_{\delta}$, i.e., all the background gradient effects are of higher order with respect to the amplitude of the fluctuations.  

In what concerns the FLR or the $\epsilon_{\perp}$- ordering, we consider models derived in the limit with full FLR corrections as well as models truncated up to the second-order in $\epsilon_{\perp}$.

 As a perturbative theory, each of the coordinate transformations,  the \textit{guiding-center} and the \textit{gyrocenter}, represents an infinite series of corrections ordered according to the corresponding small parameter, $\epsilon_B$ or $\epsilon_{\delta}$. In particular, for the particle position $\bf x$, this means that the exact gyrokinetic coordinate transformation contains an infinite series of polarization displacements.
Roughly speaking, the position of the particle $\bf x$ as a function of the reduced \textit{gyrocenter} position $\bf X$ is given by
\begin{equation*}
\mathbf{x}=\mathbf{X}+\bm\rho_0(\mathbf{X},\mu,\theta)+\bm\rho_1(\mathbf{X},\mu,\theta),
\end{equation*}
where we have introduced two polarization displacements: $\bm\rho_0$ corresponding to the lowest order \textit{guiding-center} reduction and $\bm\rho_1$ corresponding to the lowest order of the subsequent \textit{gyrocenter} reduction.

The lowest order \textit{guiding-center} displacement (in the {\em guiding-center coordinates}) is given by:
\begin{equation}
\bm\rho_{0}\equiv\frac{m c }{e}\sqrt{\frac{2 \mu}{m B}}\ \widehat{\bm\rho}\equiv\rho_0\widehat{\bm\rho}\sim\mathcal{O}(\epsilon_B^0)
\label{eq:rho_0},
\end{equation}
where $\widehat{\bm\rho}$ is the unit vector in the plane perpendicular to the background magnetic field; the magnitude of magnetic field $B$ is evaluated at the reduced ({\em guiding-center}) position $\bf X$. The general gyrokinetic derivation comes up with a result that all the following \textit{guiding-center} polarization displacements are at least of order $\mathcal{O}(\epsilon_B)$ or higher (see, for example, Eq.~(36) in \cite{Brizard_2013} or Eqs.~(63) and~(66) in \cite{Tronko_Brizard_2015}), these corrections are neglected in the numerical implementation. The lowest order \textit{gyrocenter} displacement is given by
\begin{equation}
\bm\rho_1=-\epsilon_{\delta}\frac{m c^2}{e B^2}\bm\nabla_{\perp}\left(\phi_1(\mathbf{X})-\frac{p_z}{mc}A_{1\|}(\mathbf{X})\right)\sim\mathcal{O}(\epsilon_{\delta})\mathcal{O}(\epsilon_{\perp}),
\end{equation}
where $p_z$ is the \textit{gyrocenter} scalar canonical momentum coordinate related to the parallel \textit{guiding-center} momentum and will be defined in Eq.~(\ref{eq:pz}); $\phi_1$ represents the first order perturbative electrostatic potential and $A_{1\|}$ the first order electromagnetic parallel perturbative potential.

In this work we consider the gyrokinetic coordinate transformation together with the derivation of the corresponding reduced Hamiltonian models in two cases: First, in Sec.~\ref{sec:4} we present the transformation containing all the FLR corrections, i.e., from the point of view of functional dependencies,  containing corrections of all orders related to the \textit{guiding-center} transformation where the particle is located at ${\bf x}={\bf X}+\bm\rho_0$. 
Later, in Sec.~\ref{sec:5} we explicit this change of coordinates at the lowest FLR order, which corresponds from the physical point of view to the \textit{long-wavelength} approximation, expressed in Fourier space with $\epsilon_{\perp}\equiv k_{\perp}\rho_{\mathrm{th}}\ll 1$.

\section{\label{sec:GK_pert}Phase-space perturbative procedure}

In gyrokinetic theory, the definition of the reduced phase-space coordinates is done within a common perturbative procedure together with the derivation of the reduced dynamics.
At the first step, the \textit{guiding-center} dynamical reduction starts from the local particle coordinates $\left(\mathbf{x},\mathbf{v}\right)$. To access those coordinates, one needs to define two bases of vectors: the static one and the dynamic one. The static basis is related to the background magnetic field line and the dynamic one rotates with the particle. As the static basis we choose the natural Frenet triad : the unit magnetic field vector $\widehat{\bf b}={\bf B}/B$, the normalized curvature vector 
\begin{equation*}
\widehat{\bf b}_1=\widehat{\bf b}\cdot\bm\nabla\widehat{\bf b}/|\widehat{\bf b}\cdot\bm\nabla\widehat{\bf b}|,
\end{equation*} and 
\begin{equation*}
\widehat{\bf b}_2=\widehat{\bf b}\times\widehat{\bf b}_1.
\end{equation*}
We notice that in the case of a uniform background magnetic field it is possible to choose a Cartesian frame as a static basis: $\widehat{\mathbf{b}}=\widehat{\mathbf{e}}_z, \widehat{\mathbf{b}}_1=\widehat{\mathbf{e}}_x$ and $\widehat{\mathbf{b}}_2=\widehat{\mathbf{e}}_y$.
Then, the dynamic basis $(\widehat{\bm\rho},\widehat{\mathbf{b}},\widehat{\perp})$ is defined from the static one as follows:
\begin{eqnarray}
\widehat{\bm\rho}=\widehat{\bf b}_1\cos\theta-\widehat{\bf b}_2\sin\theta, \ \
\widehat{\perp}=-\widehat{\bf b}_1\sin\theta-\widehat{\bf b}_2\cos\theta,
\label{eq:rot_bas}
\end{eqnarray}
where $\widehat{\bm\rho}$ is used for definition of the \textit{guiding-center} displacement $\bm\rho_0$ in Eq.~(\ref{eq:rho_0}) such that the local particle velocity is decomposed in the following way:
\begin{equation*}
{\bf v}= v_{\|}\widehat{\mathbf b}+\sqrt{\frac{2\mu B}{m}}\widehat{\perp}.
\end{equation*}

At the lowest order, the \textit{guiding-center} transformation is defined as follows: the position of the particle is decomposed as ${\bf x}={\bf X}+\bm\rho_0({\bf X}, \mu,\theta)$, with $\bf X$ the reduced particle (i.e., \textit{guiding-center}) position and $\bm\rho_0$ the lowest order \textit{guiding-center} polarization shift; the scalar momentum coordinate is the parallel kinetic momentum $p_{\|}=m v_{\|}$; $\mu$ is the lowest order magnetic moment given by Eq.~(\ref{eq:mu0}) and $\theta$ is the fast rotation angle.

We consider the \textit{guiding-center} phase space Lagrangian \cite{Littlejohn_1983}, \cite{Cary_Brizard} in the $\left({\bf X},p_{\|},\mu,\theta\right)$ coordinates as a starting point of the derivation:
\begin{equation}
L_{\mathrm{gc}}\left(\mathbf{X}, p_{\|},\mu,\theta\right) =\frac{e}{c}\mathbf {A}^*\cdot\dot{\mathbf X}+\frac{m c}{e}\mu\ \dot{\theta}-H_{\mathrm{gc}},
\label{gc_Lagrangian}
\end{equation}
where the symplectic part contains the modified magnetic potential:
\begin{equation*}
{\mathbf A}^*=\mathbf{A}+\frac{c}{e}\ p_{\|}\widehat{\mathbf b}.
\end{equation*}
The \textit{guiding-center} Hamiltonian is given by:
\begin{equation*}
H_{\mathrm{gc}}=\frac{p_{\|}^2}{2 m}+\mu B.
\end{equation*}

For the second step, i.e., the {\em gyrocenter} reduction, we consider the first order fluctuating time-dependent electromagnetic potential fields $\phi_1$ and $A_{1\|}$ both of order $\mathcal{O}(\epsilon_{\delta})$. Following the approximations currently performed on global code models, the perpendicular part of perturbed magnetic potential is omitted here. Moreover, as we have already stated in the introduction, the \textit{low-$\beta$} approximation is assumed in our derivation, which corresponds to the choice of considering the perpendicular component of the perturbed magnetic field only: $\mathbf{B}_1=\widehat{\mathbf b}\times\bm\nabla A_{1\|}$.  These approximations are implemented in the electromagnetic global particle-in-cell code ORB5 as well as in the global version of the Eulerian code GENE.

To account for the time-dependence of the perturbed electromagnetic potentials $A_{1\|}$ and $\phi_{1}$, we extend the phase space from $6$ to $8$ dimensions. Therefore, formally the \textit{gyrocenter}  dynamical reduction is performed on the $8$-dimensional phase space where $(t,w)$ are canonically conjugate: $t$ corresponds to time and $w$ to energy.
This extension of the phase space is used to make the dynamical system autonomous so as to perform the coordinate change in a more consistent way (see for example \cite{Arnold_1989}). From the physical point of view, the relevant reduced dynamics is still performed on the $4$ dimensional part with coordinates $\left({\bf X}, p_z\right)$, where
\begin{equation}
p_z=m v_{\|}+\epsilon_{\delta}\frac{e}{c} A_{1\|}(\mathbf{X}+\bm{\rho}_0),
\label{eq:pz}
\end{equation}
is the \textit{guiding-center} canonical momentum, and we notice that the perturbed part of magnetic potential $A_{1\|}$ enters in the definition of one of the phase space variables.
Consequently, the  perturbed \textit{guiding-center} phase-space Lagrangian and Hamiltonian are \cite{Brizard_Hahm,TBS_2016} :
\begin{eqnarray}
L_{\mathrm{pert}}\left({\mathbf X}, p_z,\mu,\theta;t,w \right)&=&
\left(\frac{e}{c}{\mathbf A}+p_z\widehat{\mathbf b}\right)\cdot\dot{\mathbf X}+\frac{mc}{e}\mu\dot{\theta}
\nonumber
\\
&-&H_{\mathrm{pert}}\left({\mathbf X}, p_z,\mu,\theta;t,w \right),
\label{eq:Lgc}
\\
H_{\mathrm{pert}}\left({\mathbf X}, p_z,\mu,\theta;t,w \right)&=&\frac{p_z^2}{2m}+\mu B+\epsilon_{\delta} e\phi_1({\mathbf X}+\bm\rho_0)
-
\epsilon_{\delta}\frac{e\ p_z}{m c}A_{1\|}({\mathbf X}+\bm\rho_0)
\nonumber
\\
&+&\epsilon_{\delta}^2\frac{1}{2m}\left(\frac{e}{c}\right)^2 A_{1\|}^2({\mathbf X}+\bm\rho_0)+w.
\label{eq:H_tilde}
\end{eqnarray}
All the background quantities $\bf A$, $\widehat{\bf b}$ and $B$ are evaluated in the reduced \textit{guiding-center} position $\bf X$,
while the perturbative electromagnetic potentials $A_{1\|}$ and $\phi_{1}$ are evaluated in the reduced particle position ${\mathbf{X}+\bm\rho_0}$, i.e., the perturbed electromagnetic potentials depend on the gyroangle through the displacement $\bm\rho_0$. 
The first two terms of the right hand side of Eq.~(\ref{eq:Lgc}) represent the non-perturbed symplectic part and $H_{\mathrm{pert}}$, given by Eq.~(\ref{eq:H_tilde}) is the perturbed Hamiltonian of the system. 

This choice of phase space coordinates corresponds to the Hamiltonian \cite{Brizard_Hahm}, or $p_z$-, \cite{Scott_Smirnov} representation of the perturbed \textit{guiding-center} dynamics. It gives the possibility of keeping the symplectic part of the phase-space Lagrangian free from all electromagnetic perturbations and therefore, gyroangle-independent.
The gyroangle-dependent terms are transferred into the expression for the perturbed Hamiltonian $H_{\mathrm{pert}}$. For this reason, the corresponding representation of the reduced dynamics is called "Hamiltonian representation". This maneuver has one significant advantage while constructing the phase-space dynamical reduction procedure for the phase-space Lagrangian: the (canonical) transformations have to be applied only on the Hamiltonian part of the phase space Lagrangian, since the corresponding symplectic part is already free of any $\theta$ dependence. 

The Hamiltonian representation is the most common choice for the models implemented in particle-in-cell simulations since it avoids the appearance of the inductive electric field (i.e., the explicit time-derivative of the perturbative magnetic potential $A_{1\|}$) in the particle characteristics given in Sec.~\ref{sec:GK_Vlasov}. For example, a control-variate scheme implemented in PIC code ORB5 \cite{Kleiber_2011} is using the Hamiltonian representation.

In the next section we show how to eliminate the gyroangle-dependence from the perturbed \textit{guiding-center} phase-space Lagrangian induced by the second-order $\sim\mathcal{O}(\epsilon_{\delta}^2)$ perturbed electromagnetic potentials.  To that purpose we build a Lie-transform near-identity transformation and move from the {\em guiding-center} to the {\em gyrocenter} variables by applying it systematically to the phase space variables and at the same time to the perturbed Hamiltonian $H_{\mathrm{pert}}$, given by Eq.~(\ref{eq:H_tilde}).
Our goal is to clarify the connection between the displacements $\bm\rho_0$ and $\bm\rho_1$, and to eliminate the gyroangle-dependence from the reduced dynamics.

\section{Full FLR Hamiltonian model}\label{sec:4}

In this section we build a near-identity phase-space transformation aiming to eliminate the gyroangle-dependence from the perturbed phase-space Lagrangian (\ref{eq:Lgc}).
Since within the $p_z$- representation the symplectic part of the phase-space Lagrangian does not contain any $\theta$-dependence, the \textit{gyrocenter} phase-space transformation will only modify its corresponding Hamiltonian part.
For the near-identity phase-space transformations we use Lie transforms which have several advantages, among them, they are canonical transformations, meaning that they do not affect the expression of the Poisson backet, only the expression of the Hamiltonian.

To define this technique, we need first of all, a Poisson bracket for the {\em guiding-center} dynamics, which can be derived from the symplectic part of the perturbed Lagrangian given by Eq.~{(\ref{eq:Lgc}), which coincides with the symplectic part of the unperturbed one defined by Eq.~(\ref{gc_Lagrangian}) (see \cite{Cary_Brizard} for more details):
\begin{eqnarray}
\{F,G\}_{\mathrm{gc}}&=&\frac{1}{\epsilon}\frac{e}{mc} \left(\frac{\partial F}{\partial \theta}\frac{\partial G}{\partial \mu}-\frac{\partial F}{\partial \mu}\frac{\partial G}{\partial \theta}\right)+\frac{{\mathbf B}^*}{B_{\|}^*}\cdot\left(\bm\nabla F\frac{\partial G}{\partial p_{z}}-\frac{\partial F}{\partial p_{z}}\bm\nabla G\right)
\nonumber
\\&-&\epsilon
\frac{c\widehat{\mathbf b}}{e B_{\|}^*}\cdot\left(\bm\nabla F\times\bm\nabla G\right)
-\frac{\partial F}{\partial w} \frac{\partial G}{\partial t}+\frac{\partial F}{\partial t}\frac{\partial G}{\partial w},
\label{GC_PB_EXT}
\end{eqnarray}
where $\mathbf{B}^{*}=\mathbf{B}+\frac{e}{c} p_z \bm\nabla\times\widehat{\mathbf b}$ and $B_{\|}^*=\mathbf{B}^{*}\cdot\widehat{\mathbf b}$. For physical reasons, we know that there is a time scale separation among the motion described by the different terms of the Poisson bracket. 
In order to exploit this separation we order the first three terms of the bracket following the formal ordering introduced by Newcomb \cite{newcomb1961lagrangian}, $\epsilon=e^{-1}$:
the first term $\propto e$ is related to the fast rotation around the magnetic field line (fastest scale of motion), the second term $\propto e^0$ represents the parallel motion, and the third term $\propto e^{-1}$ is related to the slow drifts in the perpendicular direction.
The last term in the bracket corresponds to the extension of the phase space to $8$ dimensions (autonomization of the system). 
In the following calculations we fix the small parameter $\epsilon=\epsilon_{\delta}$, so that the Poisson bracket~(\ref{GC_PB_EXT}) is separated in three brackets according to the scale of motion:
\begin{eqnarray}
\left\{F,G\right\}_{\mathrm{gc}}=\frac{1}{\epsilon_{\delta}}\left\{F,G\right\}_{-1}+\left\{F,G\right\}_{0}+\epsilon_{\delta}\left\{F,G\right\}_{1}
-\frac{\partial F}{\partial w} \frac{\partial G}{\partial t}+\frac{\partial F}{\partial t}\frac{\partial G}{\partial w},
\label{PB_scales}
\end{eqnarray}
where
\begin{eqnarray}
\label{bracket_minus_1}
\left\{F,G\right\}_{-1}&=&\frac{e}{mc} \left(\frac{\partial F}{\partial \theta}\frac{\partial G}{\partial \mu}-\frac{\partial F}{\partial \mu}\frac{\partial G}{\partial \theta}\right),
\\
\label{bracket_zero}
\left\{F,G\right\}_{0}&=&\frac{{\mathbf B}^*}{B_{\|}^*}\cdot\left(\bm\nabla F\frac{\partial G}{\partial p_{z}}-\frac{\partial F}{\partial p_{z}}\bm\nabla G\right),
\\
\label{bracket_plus_1}
\left\{F,G\right\}_{1}&=&
\frac{c\widehat{\mathbf b}}{e B_{\|}^*}\cdot\left(\bm\nabla F\times\bm\nabla G\right).
\end{eqnarray}

Following the $p_z$-representation of the Lagrangian defined in Eq.~(\ref{eq:Lgc}),
we rewrite the perturbed Hamiltonian $H_{\mathrm{pert}}$, given by Eq.~(\ref{eq:H_tilde}) as follows:
\begin{equation}
H_{\mathrm{pert}}=H_0+\epsilon_{\delta}\ e\ \psi_{1}({\mathbf{X}+\bm\rho_0},p_z)
+\epsilon_{\delta}^2\frac{1}{2 m}\left(\frac{e}{c}\right)^2 A_{1\|}^2({\mathbf{X}}+\bm\rho_0),
\label{eqn:Hgy}
\end{equation}
where the unperturbed \textit{guiding-center} Hamiltonian now writes as:
\begin{equation*}
H_0=\frac{p_{z}^2}{2 m}+\mu B+w,
\end{equation*}
and 
\begin{equation*}
\psi_{\mathrm{1}}({\bf X}+\bm\rho_0,p_z)= \phi_{1}({\mathbf{X}}+\bm\rho_0)
-\frac{p_{z}}{m c}   A_{1\|}({\mathbf{X}}+\bm\rho_0),
\end{equation*}
is the  linear perturbed \textit{gyrocenter} potential. Note that the \textit{guiding-center} displacement $\bm\rho_0$  given by Eq.~(\ref{eq:rho_0}) is depending on the phase-space coordinates $(\mathbf{X},\mu,\theta)$. To make formulas more compact we omit writing the functional dependencies of the displacement $\bm\rho_0$ explicitly when this is unambiguous. We notice that the spatial dependence of $\bm\rho_0$ can be omitted since $\left|\bm\nabla\bm\rho_0\right|\sim\mathcal{O}(\epsilon_B)$.

\subsection{\textit{Gyrocenter} phase-space coordinate transformation}

A Lie transform generated by a scalar differentiable function $S$ is defined by its action on observables $G$ as:
\begin{equation}
\bar{G}={\rm e}^{-\varepsilon\pounds_S} G= G-\varepsilon \left\{S,G\right\}+\frac{1}{2}\varepsilon^2 \left\{S,\{S,G\}\right\}+\mathcal{O}(\varepsilon^3),
\label{eqn:Lie}
\end{equation}
where $\varepsilon$ is a small parameter of a given problem and $\{\cdot,\cdot\}$ is a Poisson bracket. To this change of observables corresponds an invertible change of coordinates by the scalar invariance: $\bar{G}(\bar{\bf z})=G({\bf z})$, where the new coordinates are given by 
$$
\bar{\bf z}={\rm e}^{\varepsilon\pounds_S} {\bf z},
$$ 
where we have used the fundamental property of a Lie transform $\varphi({\rm e}^{-\varepsilon\pounds_S} G)={\rm e}^{-\varepsilon\pounds_S}\varphi(G)$ for any scalar function $\varphi$. Another fundamental property is that Lie transforms are canonical changes of coordinates, in the sense that the expression of the Poisson bracket in the new variables is exactly the same as the one in the old variables. This comes from the property $\{{\rm e}^{-\varepsilon\pounds_S} F,{\rm e}^{-\varepsilon\pounds_S} F\}={\rm e}^{-\varepsilon\pounds_S} \{F,G\}$. 

The full gyrokinetic coordinate transformation represents an infinite series of near-identity phase-space  transformations, aiming to remove the gyroangle dependence from the reduced dynamics at all orders. The key element of the reduction procedure is the identification of a generating function $S$, which plays a double role: first, defining new phase space coordinates and second, the reduced dynamics. The generating function $S$ is constructed from a perturbative series.
Keeping the chosen ordering in mind, we are using the Lie transform given by Eq.~(\ref{eqn:Lie}) with the generating function at the first order $S$, and the Poisson bracket given by Eq.~(\ref{GC_PB_EXT}), we define the \textit{gyrocenter} phase-space coordinates as:
\begin{eqnarray}
\bar{\mathbf X}&=&{\rm e}^{\epsilon_{\delta}\pounds_{S}} \mathbf{X}=\mathbf{X}+\epsilon_{\delta}\{S,\mathbf{X}\}_{\mathrm{gc}}+\mathcal{O}(\epsilon_{\delta}^2),
\label{eq:S1X}\\
\bar{p}_z&=&{\rm e}^{\epsilon_{\delta}\pounds_{S}} p_z=p_z+\epsilon_{\delta}\{S,p_z\}_{\mathrm{gc}}+\mathcal{O}(\epsilon_{\delta}^2),
\label{eq:S1pz}\\
\bar{\mu}&=&{\rm e}^{\epsilon_{\delta}\pounds_{S}}\mu=\mu+\epsilon_{\delta}\{S,\mu\}_{\mathrm{gc}}+\mathcal{O}(\epsilon_{\delta}^2),
\label{eq:S1mu}\\
\bar{\theta}&=&{\rm e}^{\epsilon_{\delta}\pounds_{S}}\theta=\theta+\epsilon_{\delta}\{S,\theta\}_{\mathrm{gc}}+\mathcal{O}(\epsilon_{\delta}^2)
\label{eq:S1theta}.
\end{eqnarray}
Therefore, in principle, there are two sets of coordinates, the {\em guiding-center} ones $(\mathbf{X},p_z,\mu,\theta)$, and the {\em gyrocenter} ones $(\bar{\mathbf{X}},\bar{p_z},\bar{\mu},\bar{\theta})$. Since, after performing the Lie transforms, it is clear that the coordinates are the {\em gyrocenter} ones, we will omit the bars over the new variables where it is not ambiguous,
and denote the \textit{gyrocenter} phase space coordinates as $(\mathbf{X},p_z,\mu,\theta)$ in what follows.

The \textit{gyrocenter} displacement is defined from the Lie transform of the position of the particle expressed in the \textit{guiding-center} coordinates:
\begin{equation}
\label{eq:epsi}
{\rm e}^{-\epsilon_{\delta}\pounds_{S}} (\mathbf{X}+\bm\rho_0)=\mathbf X+\bm\rho_0+\bm\rho_1,
\end{equation}
where
\begin{equation}
\bm\rho_1=-\epsilon_{\delta}\{S,\mathbf{X}+\bm\rho_0\}_{\mathrm{gc}},
\label{eq:rho_1}
\end{equation}
at the first order.
An explicit calculation of the polarization displacement $\bm\rho_1$ requires the knowledge of the generating function $S$, which is also necessary to get the expression for the reduced Hamiltonian and therefore to derive the reduced dynamics. This emphasizes the essential link between the definition of the new phase-space coordinates and derivation of the reduced gyroangle-independent dynamics.

We derive the expression for the reduced Hamiltonian together with the expression for the lowest order generating function $S$ in the following section.

\subsection{\label{sec:GK_Ham}Full FLR \textit{gyrocenter} dynamics}

In order to accommodate various orderings and small parameters present in the problem, and in particular to eliminate the singularity in $\epsilon_{\delta}$ appearing in the first term of the Poisson bracket (\ref{PB_scales}),
we expand the generating function $S=\epsilon_{\delta} S_1+\epsilon_{\delta}^2 S_2+\mathcal{O}(\epsilon_{\delta}^3)$. Taking  Eqns.~(\ref{PB_scales},\ref{bracket_minus_1},\ref{bracket_zero},\ref{bracket_plus_1}) into account, the corresponding Lie transform up to order $\epsilon_{\delta}^3$ is given by:
\begin{eqnarray}
{\rm e}^{-\epsilon_{\delta}\pounds_{S}}
\nonumber
&=&
\left(1-\epsilon_{\delta}^2\left\{S_1,.\right\}_{\mathrm{gc}}
+\frac{1}{2}\epsilon_{\delta}^4\left\{S_1,\left\{S_1,.\right\}_{\mathrm{gc}}\right\}_{\mathrm{gc}}
\right)
\left(1-\epsilon_{\delta}^3\left\{S_2, .\right\}_{\mathrm{gc}}\right)+\mathcal{O}(\epsilon_{\delta}^3),
\\
&=&
\left(1-\epsilon_{\delta}\{S_1,.\}_{-1}-\epsilon_{\delta}^2\left\{S_1, .\right\}_0+
\frac{1}{2}\epsilon_{\delta}^2\left\{S_1,\left\{S_1, .\right\}_{-1}\right\}_{-1}\right)
\times
\nonumber
\\
&&\left(1-\epsilon_{\delta}^2\left\{S_2, .\right\}_{-1}\right)+\mathcal{O}(\epsilon_{\delta}^3)
=1-\epsilon_{\delta}\left\{S_1, .\right\}_{-1}
\nonumber
\\
&&
-\epsilon_{\delta}^2\left\{S_1, .\right\}_0+\frac{1}{2}\epsilon_{\delta}^2\left\{S_1,\left\{S_1, .\right\}_{-1}\right\}_{-1}-\epsilon_{\delta}^2\left\{S_2, .\right\}_{-1}+\mathcal{O}(\epsilon_{\delta}^3). \label{Lie_transform}
\end{eqnarray}
The purpose of the Lie transform is to eliminate the gyroangle dependence from the reduced dynamics up to the second-order in $\epsilon_{\delta}$. As already mentioned, since the Poisson bracket (\ref{GC_PB_EXT}) is gyroangle-independent, we have to eliminate the gyroangle dependence only from the perturbed Hamiltonian (\ref{eqn:Hgy}).

Each function on the \textit{gyrocenter} phase space evaluated at the position ${\mathbf X}+\bm\rho_0$, and therefore gyroangle dependent, can be decomposed in its gyoaveraged and fluctuating parts:
\begin{equation*}
\Psi({\mathbf X}+\bm\rho_0, p_z, \mu, \theta)=\left\langle\Psi({\mathbf X}+\bm\rho_0, p_z, \mu, \theta)\right\rangle+\widetilde{\Psi}({\mathbf X}+\bm\rho_0, p_z, \mu, \theta),
\end{equation*}
where
\begin{equation*}
\left\langle\Psi({\mathbf X}+\bm\rho_0, p_z, \mu,\theta)\right\rangle=\frac{1}{2\pi}\int_0^{2\pi} \Psi({\mathbf X}+\bm\rho_0, p_z, \mu,\theta) d\theta.
\end{equation*}

First, we apply the Lie transform (\ref{Lie_transform}) to the perturbed Hamiltonian (\ref{eqn:Hgy}),  and identify a fluctuating and a gyroaveraged part for each term:
\begin{eqnarray*}
H_{\rm gy}&=&{\rm e}^{-\epsilon_{\delta}\pounds_{S}} H_{\mathrm{pert}}= H_0 +e \epsilon_{\delta}\left\langle\psi_1(\mathbf{X}+\bm\rho_0,p_z)\right\rangle+
e\epsilon_{\delta}\widetilde{\psi}_1(\mathbf{X}+\bm\rho_0,p_z)
\\
&+&
\nonumber
\epsilon_{\delta}^2\frac{1}{2m}\left(\frac{e}{c}\right)^2\left\langle A_{1\|}(\mathbf{X}+\bm\rho_0)^2\right\rangle+
\epsilon_{\delta}^2\frac{1}{2m}\left(\frac{e}{c}\right)^2\widetilde{A_{1\|}^2}(\mathbf{X}+\bm\rho_0)
\\
\nonumber
&-&
\epsilon_{\delta}\left\{S_1,H_0\right\}_{-1}-\epsilon_{\delta}^2\left\{S_1,H_0\right\}_0+
\frac{1}{2}\epsilon_{\delta}^2\left\{S_1,\left\{S_1,H_0\right\}_{-1}\right\}_{-1}
\\
\nonumber
&-&
\epsilon_{\delta}^2{\left\{S_1,e\psi_1\left(\mathbf{X}+\bm\rho_0,p_z\right)\right\}}_{-1}
-\epsilon_{\delta}^2\left\{S_2, H_0\right\}_{-1}+\mathcal{O}(\epsilon_{\delta}^3).
\end{eqnarray*}
Second, we identify order by order fluctuating contributions, which should be removed by a well-chosen generating function:
\begin{eqnarray}
\mbox{At }\mathcal{O}(\epsilon_{\delta}), \quad \ \left\{S_1,H_0\right\}_{-1}&=&e\widetilde{\psi}_1\left(\mathbf{X}+\bm\rho_0,p_z\right),
\label{order_1}
\\
\mbox{At }\mathcal{O}(\epsilon_{\delta}^2), \quad \ \left\{S_2,H_0\right\}_{-1}&=&\frac{1}{2 m}\left(\frac{e}{c}\right)^2 \widetilde{A_{1\|}^2}\left(\mathbf{X}+\bm\rho_0\right)-
{\left\{S_1,H_0\right\}}_{0}
\label{order_2}
\\
&-&
\nonumber
\oversortoftilde{\left\{S_1,e\psi_1\left(\mathbf{X}+\bm\rho_0,p_z\right)\right\}_{-1}}
+\frac{1}{2}\oversortoftilde{\left\{S_1,\left\{S_1, H_0\right\}_{-1}\right\}_{-1}}.
\end{eqnarray}

The expression for the lowest order generating function $S_1$ is obtained from Eq.~(\ref{order_1}), which represents a condition that the gyroangle-dependent part of linear electromagnetic perturbation $\widetilde{\psi}_1$ is removed from the lowest order \textit{gyrocenter} Hamiltonian.
Since $H_0$ does not depend on $\theta$, the equation determining $S_1$ reduces to:
\begin{equation}
\frac{eB}{mc} \frac{\partial S_1}{\partial\theta}=e\widetilde{\psi}_1\left(\mathbf{X}+\bm\rho_0,p_z \right).
\label{eq:S_1_bracket}
\end{equation}
Therefore, the lowest order generating function writes:
\begin{equation}
S_1=\frac{e}{\Omega}\int  d\theta\  \widetilde{\psi}_1\left(\mathbf{X}+\bm\rho_0,p_z \right).
\label{eq:S_1}
\end{equation}
We recover the conventional expression for the generating function, used for the gyrokinetic calculations in \cite{Hahm_1988}, \cite{Brizard_Hahm},\cite{Sugama_2000}.

We are now considering Eq.~(\ref{order_2}). Since $S_2$ removes the fluctuating parts from the second-order terms in the Hamiltonian, we only need to evaluate the corresponding gyroaveraged contributions to the reduced Hamiltonian.
Taking  into account Eq.~(\ref{eq:S_1_bracket}) which indicates that  $S_1$ is a purely fluctuating function,  the second term in Eq.~(\ref{order_2}), namely $ {\left\{S_1,H_0\right\}}_{0}$, is purely fluctuating and will not affect the expression of the second order reduced Hamiltonian. For similar reasons, the third term possesses the following property: $\left\langle\{S_1,\psi_1\}_{-1}\right\rangle=\left\langle\{S_1,\widetilde{\psi}_1\}_{-1}\right\rangle$, i.e., we obtain a partial cancellation of the second-order terms:
\begin{eqnarray*}
\nonumber
\frac{1}{2}\left\langle\{S_1,\{S_1,H_0\}_{-1}\}_{-1}\right\rangle
&-&
e\left\langle\{S_1,\psi_1({\mathbf X}+\bm\rho_0,p_z)\}_{-1}\right\rangle
\\
&=&
-\frac{e}{2}\left\langle\{S_1,\psi_1({\mathbf X}+\bm\rho_0,p_z)\}_{-1}\right\rangle.
\end{eqnarray*}
The next order correction to the generating function can be obtained from Eq.~(\ref{order_2}) using the expression for the generating function $S_1$ , which we have now determined in Eq.~(\ref{eq:S_1}):
\begin{eqnarray}
\label{S_1_1}
\frac{e B}{m c}\frac{\partial S_2}{\partial\theta}
&=&
\frac{1}{2m}\left(\frac{e}{c}\right)^2\widetilde{A_{1\|}^2}
-\frac{\mathbf{B}^*}{B_{\|}^*}\cdot\left(\bm\nabla S_1\frac{p_z}{m}-
\frac{\partial S_1}{\partial p_z}\mu\bm\nabla B \right)
\\
\nonumber
&-&
\frac{e^2}{2mc} \oversortoftilde{\left(\frac{\partial S_1}{\partial \theta}\frac{\partial \widetilde{\psi}_1}{\partial \mu}-\frac{\partial S_1}{\partial \mu}\frac{\partial \widetilde{\psi}_1}{\partial \theta}\right)}-
\frac{e^2}{mc} \frac{\partial S_1}{\partial \theta}\frac{\partial \langle \psi_1\rangle }{\partial \mu}.
\end{eqnarray}
Here we do not explicitly resolve Eq.~(\ref{S_1_1}), we just use the fact that $S_2$ removes all the fluctuating parts at the order $\mathcal{O}(\epsilon_{\delta}^2)$. We notice that the perturbative procedure can be expanded to higher orders.

Finally, we get the expression for the second-order Hamiltonian containing the FLR corrections (i.e., with respect to the polarization displacement $\bm\rho_0$) at all orders:

\begin{eqnarray*}
H_{\rm gy}
&=&
\frac{p_z^2}{2 m} +\mu B+w+ \epsilon_{\delta}\left(e\left\langle\phi_1 (\mathbf X+\bm\rho_0)\right\rangle\right.
\\
\nonumber
&-&\left.\frac{e}{mc}\epsilon_{\delta}\ p_z\ \left\langle A_{1\|} (\mathbf X+\bm\rho_0)\right\rangle\right)
+\epsilon_{\delta}^2
\frac{1}{2 m}\left(\frac{e}{c}\right)^2\left\langle A_{1\|}(\mathbf{X}+\bm\rho_0)^2\right\rangle
\nonumber
\\
&-&\epsilon_{\delta}^2\ \frac{e}{2}\left\langle\{S_1,\widetilde{\psi}_1({\mathbf X}+\bm\rho_0,p_z)\}_{-1}\right\rangle,
\nonumber
\end{eqnarray*}
where $S_1$ is given by Eq.~(\ref{eq:S_1}).
Rewriting the expression for the lowest order Poisson bracket:
\begin{equation*}
\left\{F,G\right\}_{-1}=\frac{e}{mc}\frac{\partial}{\partial\theta}\left(F \frac{\partial G}{\partial\mu}\right)-\frac{e}{mc}\frac{\partial}{\partial\mu}\left(F\frac{\partial G}{\partial\theta}\right),
\end{equation*}
we have
\begin{eqnarray*}
\{S_1,\widetilde{\psi}_1(\mathbf{X}+\bm\rho_0,p_z)\}_{-1}&=&
\frac{e}{mc}\frac{\partial}{\partial\mu}\left(\widetilde{\psi}_1(\mathbf{X}+\bm\rho_0,p_z)\frac{\partial S_1}{\partial\theta}\right)
\\
&-&
\frac{e}{mc}\frac{\partial}{\partial\theta}\left(\widetilde{\psi}_1(\mathbf{X}+\bm\rho_0,p_z)\frac{\partial S_1}{\partial\mu}\right),
\end{eqnarray*}
and taking  into account that $\partial_{\theta} S_1=\frac{e}{\Omega}\widetilde{\psi}_1(\mathbf{X}+\bm\rho_0,p_z)$, we get:
\begin{eqnarray}
\left\langle\{S_1,\widetilde{\psi}_1(\mathbf{X}+\bm\rho_0,p_z)\}_{-1}\right\rangle=\frac{e}{B} \partial_{\mu}\left\langle\widetilde{\psi}_1\left(\mathbf{X}+\bm\rho_0,p_z\right)^2\right\rangle.
\label{eq:FullFLR}
\end{eqnarray}
Therefore, the reduced Hamiltonian becomes:
\begin{eqnarray}
H_{\rm gy}
\nonumber
&=&
\frac{p_z^2}{2 m} +\mu B+w+ \epsilon_{\delta}\left(e \left\langle\phi_1 (\mathbf X+\bm\rho_0)\right\rangle\right.
\\
\nonumber
&-&\left.\frac{e}{mc}\ p_z\ \left\langle A_{1\|} (\mathbf X+\bm\rho_0)\right\rangle\right)
+\epsilon_{\delta}^2
\frac{1}{2 m} \left(\frac{e}{c}\right)^2 \left\langle A_{1\|}(\mathbf X+\bm\rho_0)^2\right\rangle
\\
&-&\epsilon_{\delta}^2\frac{e^2}{2 B}\partial_{\mu}\left\langle\widetilde{\psi}_1\left(\mathbf{X}+\bm\rho_0,p_z\right)^2\right\rangle.
\label{eq:H_2_full}
\end{eqnarray}

For further convenience, we decompose the expression for the Hamiltonian as $H_{\rm gy}=\overline{H}_0+\epsilon_{\delta}\overline{H}_1+\epsilon_{\delta}^2\overline{H}_{2}$ with:
\begin{eqnarray}
\overline{H}_0&=&\frac{p_z^2}{2 m} +\mu B+w
\label{H_0}
\\
\overline{H}_1&=&e\ \left\langle\phi_1 (\mathbf X+\bm\rho_0)\right\rangle
-\frac{e}{mc}\ p_z\ \left\langle A_{1\|} (\mathbf X+\bm\rho_0)\right\rangle
\label{H_1}
\\
\overline{H}_{2}&=&
\frac{1}{2 m} \left(\frac{e}{c}\right)^2 \left\langle A_{1\|}(\mathbf X+\bm\rho_0)^2\right\rangle
-\frac{e^2}{2 B}\partial_{\mu}\left\langle\widetilde{\psi}_1\left(\mathbf{X}+\bm\rho_0,p_z\right)^2\right\rangle,
\label{H_2full}
\end{eqnarray}
where $\overline{H}_0$ is the non-perturbed \textit{guiding-center} Hamiltonian, $\overline{H}_1$ is the first order \textit{gyrocenter} Hamiltonian and $\overline{H}_2$ is the second-order \textit{gyrocenter} Hamiltonian, which contains the FLR corrections at all orders.
This result corresponds to the expression obtained in a slab magnetic geometry \cite{Hahm_Lee_1988}.

We recall that since the expression for the generating function $S$ has been defined within the dynamical reduction procedure for the Hamiltonian, we have obtained the explicit expression for the second-order Hamiltonian without writing explicitly the expression for the new (i.e., {\em gyrocenter}) phase space coordinates. We are now completing the dynamical reduction procedure by providing an explicit expression for the \textit{gyrocenter} phase space coordinates.

\subsection{Explicit \textit{gyrocenter} phase space coordinate transformation}
Using the expression for the generating function $S_1$ given by Eq.~(\ref{eq:S_1}) and the \textit{guiding-center} Poisson bracket given by Eq.~(\ref{GC_PB_EXT}), we explicitly evaluate the expression for the corresponding lowest order \textit{gyrocenter} coordinate transformation given by Eqs.~(\ref{eq:S1X}-\ref{eq:S1theta}), i.e., we express the leading order modifications of the {\em gyrocenter} coordinates $(\bar{\bf X},\bar{p_z},\bar{\mu},\bar{\theta})$ as functions of the {\em guiding-center} coordinates $({\bf X},p_z,\mu,\theta)$:
\begin{eqnarray*}
&& \bar{\bf X}={\bf X}-\epsilon_\delta^2 \frac{{\bf B}^*}{B^*_\parallel B}\int d\theta \widetilde{A_{1\|}}({\bf X}+\bm\rho_0)+O(\epsilon_\delta^3),\\
&& \bar{p_z}=p_z+\epsilon_\delta^2\frac{e}{\Omega}\int d\theta \frac{{\bf B}^*}{B^*_\parallel B}\cdot \nabla \widetilde{\psi}_1({\bf X}+\bm\rho_0,p_z)+O(\epsilon_\delta^3),\\
&& \bar{\mu}=\mu+\epsilon_\delta \frac{e}{B}\widetilde{\psi}_1({\bf X}+\bm\rho_0,p_z)+O(\epsilon_\delta^2),\\
&& \bar{\theta}=\theta-\epsilon_\delta \frac{e}{B}\int d\theta \frac{\partial \widetilde{\psi}_1}{\partial \mu}({\bf X}+\bm\rho_0,p_z)+O(\epsilon_\delta^2).
\end{eqnarray*}
We notice that the modifications for $\bar{\bf X}$ and $\bar{p_z}$ are of order $\epsilon_\delta^2$, whereas the modifications for $\bar{\mu}$ and $\bar{\theta}$ are of order $\epsilon_\delta$. Because of the modification of $\bar{\mu}$ and $\bar{\theta}$ at the order $\epsilon_\delta$, the {\em gyrocenter} displacement is of order $\epsilon_\delta$. 

Concerning the coordinate change, we recall that there are two important sets of coordinates in the two-step gyrokinetic reduction, the {\em guiding-center} coordinates $({\bf X},p_z,\mu,\theta)$ and the {\em gyrocenter} ones  $(\bar{\bf X},\bar{p_z},\bar{\mu},\bar{\theta})$ which we have also denoted $({\bf X},p_z,\mu,\theta)$ after Lie transforming the Hamiltonian. The position of the particle is ${\bf x}={\bf X}+{\bm \rho}_0$ in the {\em guiding-center} coordinates and ${\bf x}={\bf X}+{\bm \rho}_0+{\bm \rho}_1$ in the {\em gyrocenter} coordinates (which, for clarity, should be denoted ${\bf x}=\bar{\bf X}+\bar{\bm \rho}_0+\bar{\bm \rho}_1$). Therefore, in the {\em gyrocenter} coordinates $ {\bm \rho}_0+{\bm \rho}_1$ is the difference between the position of the particle and the position of the {\em gyrocenter}, whereas $\bm\rho_0$ is the difference between the position of the particle and the {\em guiding center} in the {\em guiding-center} coordinates. It is tempting to conclude that the {\em gyrocenter} displacement ${\bm \rho}_1$ is the distance between the {\em guiding center} and the {\em gyrocenter}. However this is incorrect. We have seen above that the {\em gyrocenter} displacement is of order $\epsilon_\delta$ whereas the distance between the {\em guiding center} and the {\em gyrocenter} is of order $\epsilon_\delta^2$. The way to solve the apparent contradiction is to carefully evaluate the system in which the positions are expressed. The expressions of ${\bm \rho}_0$ in the {\em guiding center} and in the {\em gyrocenter} coordinates are different, even at the leading order: At the leading order, the difference between these two expressions is given by
$$
{\bm \rho}_0({\bf X},\mu,\theta)-{\bm \rho}_0(\bar{\bf X},\bar{\mu},\bar{\theta})\approx -\epsilon_\delta \{S_1,{\bm \rho}_0\}_{-1},
$$
where we have to distinguish both sets of coordinates since they are used in the same equation. The shift in ${\bm \rho}_0$ by the Lie transform generates exactly the {\em gyrocenter} displacement ${\bm \rho}_1$ at the leading order, whereas the reduced positions remain approximately unchanged (at least at the order $\epsilon_\delta$). 

To account for all FLR corrections, i.e., corrections related to the displacement $\bm\rho_0$, in the expressions for the reduced Hamiltonian given by Eqs.~(\ref{H_0}-\ref{H_2full}) we have evaluated the fields $\phi_1$ and $A_{1\|}$ at the position $\bar{\mathbf X}+\bm{\rho}_0(\bar{\mathbf X},\bar{\mu},\bar{\theta})$ instead of the particle position $\mathbf{x}=\bar{\mathbf X}+\bm{\rho}_0(\bar{\mathbf X},\bar{\mu},\bar{\theta})+\bm{\rho}_1(\bar{\mathbf X},\bar{\mu},\bar{\theta})$. This generates polarization and magnetization corrections related to the displacement  $\bm\rho_1$ into the second term of the second order Hamiltonian $\bar{H}_2$ given by Eq.~(\ref{H_2full}).
To make it more explicit, we consider the Lie transform of the {\em gyrocenter} potential $\psi_1$ evaluated at the position $(\mathbf{X}+\bm\rho_0({\mathbf X},\mu,\theta))$:
\begin{equation*}
\left. {\rm e}^{-\epsilon_{\delta}\pounds_{S}} \psi_1({\mathbf X}+\bm\rho_0({\mathbf X},\mu,\theta),p_z)\right\vert_{(\bar{\bf X},\bar{p_z},\bar{\mu},\bar{\theta})}=\psi_1(\bar{\mathbf X}+\bm{\rho}_0(\bar{\mathbf X},\bar{\mu},\bar{\theta})+\bm{\rho}_1(\bar{\mathbf X},\bar{\mu},\bar{\theta}),\bar{p_z}),
\end{equation*}
which is obtained from Eq.~(\ref{eq:epsi}). 
The expansion of the left-hand side leads to
\begin{equation*}
\left. {\rm e}^{-\epsilon_{\delta}\pounds_{S}} \psi_1({\mathbf X}+\bm\rho_0,p_z)\right\vert_{(\bar{\bf X},\bar{p_z},\bar{\mu},\bar{\theta})}=
 \psi_1(\bar{\mathbf X}+\bar{\bm\rho_0},\bar{p_z})-\epsilon_{\delta}\left\{S_1,\psi_1\right\}_{-1}+\mathcal{O}(\epsilon_{\delta}^2),
\end{equation*}
where $\bar{\bm\rho_0}=\bm{\rho}_0(\bar{\mathbf X},\bar{\mu},\bar{\theta})$.
The Taylor expansion of the right-hand side leads to , since :
\begin{equation*}
\psi_1(\bar{\mathbf X}+\bar{\bm{\rho}_0}+\bar{\bm{\rho}_1},p_z)=
\psi_1(\bar{\mathbf X}+\bar{\bm{\rho}_0},\bar{p_z})+\epsilon_{\delta} \bar{\bm{\rho}_1}\cdot \bm\nabla\psi_1
+\mathcal{O}(\epsilon_{\delta}^2),
\end{equation*}
since $\bm\rho_1\propto\mathcal{O}(\epsilon_{\delta})$.
Therefore we have $\left\{S_1,\psi_1\right\}_{-1}=-\bm\rho_1\cdot\bm\nabla\psi_1$ evaluated at $\bar{\mathbf X}+\bar{\bm\rho_0}$. In other words, according to Eqs.~(\ref{eq:FullFLR}) and (\ref{H_2full}), the polarization effects associated with the displacement $\bm\rho_1$ are contained in the second order {\em gyrocenter} Hamiltonian $\bar{H}_2$.

\section{Hamiltonian model in \textit{long-wavelength} approximation}
\label{sec:5}

In the previous section, we have derived the \textit{gyrocenter} Hamiltonian model (\ref{eq:H_2_full}) containing up to $\epsilon_{\delta}^2$ corrections.
In this section we consider an additional ordering $\epsilon_{\perp}= k_{\perp}\rho_{\mathrm{th}}\ll 1$ and derive the Hamiltonian model in the \textit{long-wavelength} approximation, containing up to the second-order FLR corrections, i.e., terms of order $\mathcal{O}(\epsilon_{\perp}^2)$.
Such a derivation has a direct interest with respect to the numerical implementation of gyrokinetic Vlasov-Maxwell models. Indeed, from the point of view of the variational derivation, the second-order terms in Eq.~(\ref{eq:H_2_full}) will enter the gyrokinetic Maxwell equations as the linear (i.e., of order $\mathcal{O}(\epsilon_{\delta})$) polarization corrections (see for example a detailed derivation in \cite{TBS_2016}).

 The gyrokinetic Vlasov-Poisson model with linearized polarization term derived in the \textit{long-wavelength} limit with $\epsilon_{\perp}= k_{\perp}\rho_{\mathrm{th}}\ll 1$ has been the first implemented model in a gyrokinetic PIC code \cite{Lee_1983}. It is still widely used for the investigation of MHD modes, and it is also suitable for studies of turbulence generated by the interaction of modes with low toroidal numbers. As it has been shown in the latest linear electromagnetic benchmark \cite{Goerler_Tronko_2016}, the \textit{long-wavelength} approximation implemented into the ORB5 code allows one to treat modes with $\epsilon_{\perp}=k_{\perp}\rho_{\mathrm{th}}< 0.6$.  

From the numerical point of view, the \textit{long-wavelength} approximate model, containing up to $\mathcal{O}(\epsilon_{\perp}^2)$ corrections is of special interest because of its consistency with respect to the $n$-point gyroaverage operator approximation, implemented in ORB5. The $n$-point operator is an extension of the original four-point gyroaverage proposed in \cite{Lee_1983}.

The expression for the gyroaverage operator, applied to a scalar field $\psi(\mathbf{X}+\bm\rho_0,p_z,\mu,\theta)$ can be expressed in Fourier space in the following form:
\begin{eqnarray*}
\left(\mathcal{J}_0^{\mathrm{gc}}\psi \right)(\mathbf{X},p_z,\mu)&=&\frac{1}{2\pi}\int_0^{2\pi}\psi\left(\mathbf{X}+\bm\rho_0(\mathbf{X},\mu,\theta),p_z,\mu\right)d\theta,
\\
\nonumber
&=&\frac{1}{(2\pi)^3}\int \widehat{\psi}(\mathbf{k}) \mathrm{J}_0(k_{\perp}\rho_{\mathrm{th}}){\rm e}^{i{\mathbf k}\cdot{\mathbf X}} d{\mathbf k},
\end{eqnarray*}
where $\widehat{\psi}$ is the Fourier-transformed scalar field $\psi$. The last expression shows that the action of the operator $\mathcal{J}_0^{\mathrm{gc}}$ in Fourier space is translated into the multiplication of  the Fourier coefficients by the Bessel function $\mathrm{J}_0(k_{\perp}\rho_i)$. In order to be implementable, the gyroaverage procedure is approximated by an average over a finite number of points on the gyroring. In the first PIC code \cite{Lee_1983} the four-point gyroaverage was used, this procedure is equivalent to considering the finite difference approximation of the Taylor expansion $\mathrm{J}_0(k_{\perp}\rho_i)\approx 1-(k_{\perp}\rho_i)^2/4$, which corresponds in real space to:
\begin{equation*}
(\mathcal{J}_0^{\mathrm{gc}}\psi)({\mathbf X},p_z,\mu)\approx \psi({\mathbf X},p_z,\mu)+
\frac{1}{4}\rho_i^2\bm\nabla_{\perp}^2\psi({\mathbf X},p_z,\mu).
\end{equation*}
A detailed discussion is available in \cite{Bottino_Sonnendruecker}.
 
Recently, a full FLR solver for the Poisson equation has been implemented in ORB5 \cite{PhD_Dominsky_2016}, \cite{Dominski_2017}. Preliminary results show that the new algorithm is twice slower than the \textit{long-wavelength} solver. This is mostly due to the need for additional integration points for the gyroaverage algorithm as it was already shown in \cite{Mishchenko_Koenies_2005}.

We present two different ways to obtain the simplified Hamiltonian model, containing up to the $\epsilon_{\perp}^2$ terms.
First of all, we perform the lowest order FLR series truncation directly on the perturbative electromagnetic potential $\widetilde{\psi}_1$ and we keep up to the second-order FLR terms inside the magnetic potential $A_{1||}$, which enter into the expression of the second-order $\mathcal{O}(\epsilon_{\delta}^2)$ electromagnetic potential given by Eq.~(\ref{eq:FullFLR}). 
Next, we are following the main steps of the \textit{gyrocenter} coordinate transformation by introducing the FLR truncation at each step: starting with the expression for the generating function $S_1$,  getting the corresponding \textit{gyrocenter} change of coordinates and finally the expression for the simplified Hamiltonian. 

In this section, we follow the direct $\epsilon_{\perp}$- truncation of Eq.~$(\ref{eq:FullFLR})$ and in Sec.~{\ref{sec:FLR_coord}} we detail the main steps of the second derivation, in which we show that the first order $\epsilon_{\perp}$-truncation of the generating function $S_1$ is sufficient for recovering the second-order FLR, i.e., $\epsilon_{\perp}^2$ corrections in the Hamiltonian.

\subsection{\label{subsec:direct}Direct full FLR model truncation}

Here we evaluate the second-order, i.e., $\mathcal{O}(\epsilon_{\perp}^2)$, FLR contributions to the second-order Hamiltonian (\ref{H_2full}).
We start by decomposing the first-order fluctuating electromagnetic potential in the FLR series:
\begin{equation*}
\widetilde{\psi}_1(\mathbf{X}+\bm\rho_0,p_z)=\bm\rho_0\cdot\bm\nabla\psi_1(\mathbf{X},p_z) +\oversortoftilde{\bm\rho_0\bm\rho_0 : \bm\nabla\bm\nabla\psi_1(\mathbf{X},p_z)}+\mathcal{O}(\epsilon_{\perp}^3).
\end{equation*}
To get the $\epsilon_{\perp}^2$- contributions in the Hamiltonian, we only keep the first term, and we calculate:
\begin{eqnarray*}
&&\frac{\partial}{\partial\mu}\left\langle\widetilde{\psi}_1\left(\mathbf{X}+{\bm\rho}_0(\mu,\theta),p_z\right)^2\right\rangle
=
\frac{\partial}{\partial\mu}\left\langle\left|\bm\rho_0\cdot\bm\nabla\psi_1(\mathbf{X},p_z)\right|^2\right\rangle=
\\
\nonumber
&=&
\frac{\partial \rho_0^2}{\partial\mu}\left\langle\widehat{\bm\rho}\widehat{\bm\rho}:\bm\nabla\psi_1(\mathbf{X},p_z)\bm\nabla\psi_1(\mathbf{X},p_z)\right\rangle
=\left(\frac{c}{e}\right)^2\frac{m}{B} \left|\bm\nabla_{\perp}\psi_1(\mathbf{X},p_z)\right|^2,
\end{eqnarray*}
where we have used the definition (\ref{eq:rho_0}), the fact that $\frac{1}{2}\frac{\partial\rho_0^2}{\partial\mu}=\frac{c^2 m}{e^2 B}$ and the dyadic tensors property $\left\langle\widehat{\bm\rho}\widehat{\bm\rho}\right\rangle=\frac{1}{2}\left(\widehat{\mathbf{b}}_1\widehat{\mathbf{b}}_1+\widehat{\mathbf{b}}_2\widehat{\mathbf{b}}_2\right)=\frac{1}{2}\mathbf{1}_{\perp}$.

The magnetic term of the second-order is truncated up to the second FLR correction:
\begin{eqnarray*}
&&\left\langle A_{1\|}(\mathbf{X}+\bm\rho_0)^2\right\rangle=
\left\langle \left( A_{1\|}(\mathbf{X})
+
{\bm\rho}_{0}\cdot\bm\nabla A_{1\|}(\mathbf{X})
+\frac{1}{2}{\bm\rho}_0{\bm\rho}_0:\bm\nabla\bm\nabla A_{1\|}(\mathbf{X})\right)^2\right\rangle
\nonumber
\\
&=&A_{1\|}^2(\mathbf{X})+m \left(\frac{c}{e}\right)^2\ \frac{\mu}{B}\left\vert\bm\nabla_{\perp}A_{1\|}(\mathbf{X})\right\vert^2+
m \left(\frac{c}{e}\right)^2\ \frac{\mu}{B}A_{1\|}(\mathbf{X}) \  \bm\nabla_{\perp}^2 A_{1\|}(\mathbf{X}).
\end{eqnarray*}
We note that the second term is missing in the ORB5 model \cite{TBS_2016}, which corresponds to the slab geometry result obtained in \cite{Hahm_Lee_1988}.
Finally, truncated up to the second-order, the FLR Hamiltonian writes:
\begin{eqnarray}
H^{\mathrm{FLR}}_{\rm gy}
\nonumber
&=&
\frac{p_z^2}{2 m} +\mu B+w+ \epsilon_{\delta}\left(e\ \left\langle\phi_1 (\mathbf X+\bm\rho_0)\right\rangle
-\frac{e}{mc}\ p_z\ \left\langle A_{1\|} (\mathbf X+\bm\rho_0)\right\rangle\right)
\label{eq:H_FLR}
\\
&+&\epsilon_{\delta}^2
\left(
\frac{1}{2 m} \left(\frac{e}{c}\right)^2A_{1\|}^2(\mathbf{X})
+ m \left(\frac{c}{e}\right)^2\ \frac{\mu}{B}\left\vert\bm\nabla_{\perp}A_{1\|}(\mathbf{X})\right\vert^2
\right)
\\
&+&
\epsilon_{\delta}^2
\left(
m \left(\frac{c}{e}\right)^2\ \frac{\mu}{B}A_{1\|}(\mathbf{X}) \  \bm\nabla_{\perp}^2 A_{1\|}(\mathbf{X})
\nonumber
-
\frac{m c^2}{2 B^2}\left| \bm\nabla_{\perp} \phi_1(\mathbf X)-\frac{1}{m c}\ p_z\bm\nabla_{\perp}A_{1 \|} (\mathbf X) \right|^2
\right),
\nonumber
\end{eqnarray}
which for further convenience is separated into three parts: the unperturbed part $\bar{H}_0$ and the first order contribution $\bar{H}_1$ are defined by Eqs.~(\ref{H_0}) and (\ref{H_1}), while the second-order term is defined as:
\begin{eqnarray}
\overline{H}_2^{\mathrm{FLR}}&=&
\label{H_2_FLR}
\frac{1}{2 m} \left(\frac{e}{c}\right)^2A_{1\|}^2(\mathbf{X})
+ m \left(\frac{c}{e}\right)^2\ \frac{\mu}{B}\left\vert\bm\nabla_{\perp}A_{1\|}(\mathbf{X})\right\vert^2
\\
&+&
m \left(\frac{c}{e}\right)^2\ \frac{\mu}{B}A_{1\|}(\mathbf{X}) \  \bm\nabla_{\perp}^2 A_{1\|}(\mathbf{X})
\nonumber
-
\frac{m c^2}{2 B^2}\left| \bm\nabla_{\perp} \phi_1(\mathbf X)-\frac{1}{mc}\ p_z\bm\nabla_{\perp}A_{1 \|} (\mathbf X) \right|^2.
\end{eqnarray}
\subsection{\label{sec:FLR_coord}\textit{Gyrocenter} coordinate transformation in the \textit{long-wavelength} approximation}

In this section we derive the truncated  Hamiltonian model (\ref{eq:H_FLR}) by performing the \textit{gyrocenter} coordinate transformation (\ref{eq:S1X}-\ref{eq:S1theta}) at the lowest FLR order. To that purpose, we only take  into account the lowest order FLR correction to the generating function $S=\epsilon_{\delta}S_1+\mathcal{O}(\epsilon_{\delta}^2)$. 
Since the lowest order FLR correction to  the fluctuating electromagnetic potential is $\widetilde{\psi}_1\left(\mathbf{X}+\bm\rho_0,p_z\right)=\rho_0\widehat{\bm\rho}\cdot\bm\nabla\psi_1(\mathbf{X},p_z)$, from Eq.~(\ref{eq:S_1}) using the property of the rotating basis vectors $\widehat{\perp}=-\int d\theta\ \widehat{\bm\rho}$ , we get:  
\begin{equation*}
S_1=-\epsilon_{\delta}\frac{mc}{B}\rho_0\ \widehat{\perp}\cdot\bm\nabla\psi_1(\mathbf{X}).
\end{equation*}
We calculate the \textit{gyrocenter} displacement $\bm\rho_1$ using the definition (\ref{eq:rho_1}),
taking  into account the lowest order Poisson bracket (\ref{bracket_minus_1}):
\begin{eqnarray*}
\bm\rho_1=-\epsilon_{\delta} \left\{S_1,{\bf X}+\bm\rho_0\right\}_{-1}=-\epsilon_{\delta}\frac{e}{mc}\left(\frac{\partial S_1}{\partial\theta}\frac{\partial\bm\rho_0}{\partial\mu}-\frac{\partial\bm\rho_0}{\partial\theta}\frac{\partial S_1}{\partial\mu}\right).
\end{eqnarray*}
From the definition of rotating basis vectors (\ref{eq:rot_bas}), and $\partial_{\mu}\rho_0^2=\frac{2 m c^2}{e^2 B}$, we have:
\begin{eqnarray*}
\frac{e}{mc}\frac{\partial S_1}{\partial\theta}\frac{\partial\bm\rho_0}{\partial\mu}=\frac{m c^2}{e B^2}\ \widehat{\bm\rho}\widehat{\bm\rho}\cdot\bm\nabla\psi_1 \ \ \mathrm{and} \
-\frac{e}{mc}\frac{\partial\bm\rho_0}{\partial\theta}\frac{\partial S_1}{\partial\mu}=\frac{m c^2}{e B^2}\ \widehat{\perp}\widehat{\perp}\cdot\bm\nabla\psi_1.
\end{eqnarray*}

Using the definition of the dyadic tensor $\bf{1}_{\perp}\equiv\widehat{\bm\rho}\widehat{\bm\rho}+\widehat{\perp}\widehat{\perp}$, the first order \textit{gyrocenter} displacement in the \textit{long wavelength approximation} is:
\begin{equation*}
\bm\rho_1\approx-\epsilon_{\delta}\frac{m c^2}{e B^2}\bm\nabla_{\perp}\psi_1.
\end{equation*}

At the leading order, only the {\em gyrocenter} phase-space coordinates $(\mu,\theta)$ are transformed according to the Eqs.~(\ref{eq:S1mu},\ref{eq:S1theta}) using:
\begin{eqnarray*}
\left\{S_1,\mu\right\}_{-1}&=&\frac{e}{mc}\frac{\partial S_1}{\partial\theta}=\frac{e}{B}\rho_0\widehat{\bm\rho}\cdot\bm\nabla\psi_1(\mathbf{X},p_z),\\
\left\{S_1,\theta\right\}_{-1}&=&-\frac{e}{mc}\frac{\partial S_1}{\partial\mu}=\frac{e}{B}\left(\partial_{\mu}\rho_0\right)\widehat{\perp}\cdot\bm\nabla\psi_1(\mathbf{X},p_z),
\end{eqnarray*}
where we have used some properties of the rotating basis vectors: $\partial_{\theta}\widehat{\perp}=-\widehat{\bm\rho}$ and 
$\partial_{\theta}\widehat{\bm\rho}=\widehat{\perp}$.

Therefore, truncating the second order {\em gyrocenter} Hamiltonian (obtained by Lie transform generated by $S_1$ containing all the FLR corrections) up to the second order in FLR corrections is equivalent to the second order {\em gyrocenter} Hamiltonian obtained from the Lie transform generated by the first FLR correction of the generating function $S_1$.

\section{\label{sec:GK_Vlasov}Gyrokinetic Vlasov equation}

In this section we derive an expression for the gyrokinetic Vlasov equation corresponding to the Hamiltonian characteristics according to the models obtained in Secs.~\ref{sec:4} and \ref{sec:5}.

The particle characteristics are calculated as follows:
\begin{eqnarray}
\label{char_X}
\dot{\mathbf X}=\left\{{\mathbf{X}},H_{\rm gy}\right\}_{\mathrm{gc}}&=&\frac{c\widehat{\mathbf b}}
{e B_{\|}^{*}}\times\bm\nabla H_{\rm gy}+\frac{\partial H_{\rm gy}}{\partial{{p}_{z}}}\frac{\mathbf{B}^{*}}{B_{\|}^{*}},\\
\label{char_p_z}
\dot{p}_{z}=\left\{p_z,H_{\rm gy}\right\}_{\mathrm{gc}}&=&-\frac{\mathbf{B}^{*}}{B_{\|}^{*}}\cdot \bm\nabla H_{\rm gy},
\end{eqnarray}
where $H_{\rm gy}$ is the Hamiltonian chosen according to the approximation, full FLR model (\ref{eq:H_2_full}) or truncated up to second FLR order model (\ref{eq:H_FLR}). The Poisson bracket is given by Eq.~(\ref{GC_PB_EXT}).
Then the corresponding Vlasov equation suitable for the $\delta f$ simulations, i.e., including the decomposition $F=F_0+\epsilon_{\delta} F_1$ into the background $F_0$ and the fluctuating part $F_1$ of the particle distribution function, is reconstructed from the characteristics given by Eqs.~(\ref{char_X}) and (\ref{char_p_z}) by saying that the distribution function is constant along the characteristics:
\begin{eqnarray*}
0=\frac{d (F_0+\epsilon_{\delta}F_1)}{dt}&=&\frac{\partial  (F_0+\epsilon_{\delta}F_1)}{\partial t}+\{{\mathbf X},H\}_{\mathrm{gc}}\cdot\bm\nabla \left(F_0+\epsilon_{\delta} F_1\right)\nonumber \\
&& \quad+\{p_z,H\}_{\mathrm{gc}}\ \partial_{p_z} \left(F_0+\epsilon_{\delta} F_1\right),
\end{eqnarray*}
and then
\begin{equation}
\label{Vlasov_bracket}
\frac{\partial  (F_0+\epsilon_{\delta}F_1)}{\partial t}=-\{F_0+\epsilon_{\delta}F_1,H\}_{\mathrm{gc}},
\end{equation}
where we have taken into account the functional dependencies of the \textit{gyrocenter} distribution function $F=F({\mathbf{X}},p_z,\mu)$ and the fact that $\mu$ is constant on the reduced {\em gyrocenter} phase space.
We assume that $\left\{F_0,\bar{H}_0\right\}_{\mathrm{gc}}=0$, i.e., the background particle distribution is time-independent; therefore there are two linear and two non-linear terms).
Following the Hamiltonian decomposition given by Eqs.~(\ref{H_0}), (\ref{H_1}), the Vlasov equation (\ref{Vlasov_bracket}) can further be expanded as:
\begin{equation*}
\frac{\partial  F_1}{\partial t}=-\left\{F_0,\bar{H}_1\right\}_{\mathrm{gc}}-\left\{F_1,\bar{H}_0\right\}_{\mathrm{gc}}-
\epsilon_{\delta}\left\{F_1,\bar{H}_1\right\}_{\mathrm{gc}} -\epsilon_{\delta}\left\{F_0,\bar{H}_2\right\}_{\mathrm{gc}},
\end{equation*}
where $\bar{H}_2$ is the second-order Hamiltonian chosen according to the approximation, full FLR model (\ref{H_2full}) or FLR truncated model (\ref{H_2_FLR}). The last term refers to the evolution of the background distribution under the second-order Hamiltonian. This term is systematically neglected in most of global codes, but is necessary for full-$f$ nonlinear simulations.

\section{Conclusions}

We have performed a derivation of the second-order (in $\epsilon_{\delta}$) \textit{gyrocenter} Hamiltonian models in the case with full FLR corrections and then considering the \textit{long-wavelength} approximation. The \textit{long-wavelength} model has been obtained in two different ways: by using a direct truncation of the 
full FLR model and by constructing the dynamical reduction procedure with the \textit{gyrocenter} generating function, containing only the lowest order FLR contributions. Whether the FLR truncation is performed a priori on the generating function, or a posteriori on the reduced Hamiltonian does not change the final expressions at the leading orders. In the course of the reduction, we have clarified the origin of the {\em gyrocenter} displacement, responsible for polarization and magnetization terms in the reduced gyrokinetic Vlasov-Maxwell equations, using a proper ordering of the generating function of the Lie transform.

\section*{Acknowledgments}
This work has been carried out within the framework of
the EUROfusion Consortium and has received funding from
the Euratom research and training Programme No. 2014-
2018 under Grant Agreement No. 633053. The views and
opinions expressed herein do not necessarily reflect those of
the European Commission.


\begin{thebibliography}{10}

\bibitem{Frieman_Chen_1982}
E.~A. {Frieman} and L.~{Chen}.
\newblock {Nonlinear gyrokinetic equations for low-frequency electromagnetic
  waves in general plasma equilibria}.
\newblock {\em Physics of Fluids}, 25(3):502--508, 1982.

\bibitem{Littlejohn_1983}
R.~G. {Littlejohn}.
\newblock {Variational principles of guiding centre motion}.
\newblock {\em {Journal of Plasma Physics}}, 29(FEB):111--125, 1983.

\bibitem{Brizard_Hahm}
A.~J. {Brizard} and T.~S. {Hahm}.
\newblock {Foundations of nonlinear gyrokinetic theory}.
\newblock {\em Reviews of Modern Physics}, 79(2):421--468, 2007.

\bibitem{Garbet_Idomura_2010}
X.~{Garbet}, Y.~{Idomura}, L.~{Villard}, and T.~H. {Watanabe}.
\newblock {Gyrokinetic simulations of turbulent transport}.
\newblock {\em {Nuclear Fusion}}, 50(4):043002, 2010.

\bibitem{TBG_2017}
N.~{Tronko}, T.~{Bottino}, A.~{Goerler}, E.~{Sonnendr\"ucker}, D.~{Told}, and
  L.~{Villard}.
\newblock {Verification of Gyrokinetic codes: theoretical background and
  applications.}
\newblock {\em {submitted to Physics of Plasmas}}.

\bibitem{Jolliet_2007}
S.~{Jolliet}, A.~{Bottino}, P.~{Angelino}, R.~{Hatzky}, T.~M. {Tran}, B.~F.
  {Mcmillan}, O.~{Sauter}, K.~{Appert}, Y.~{Idomura}, and L.~{Villard}.
\newblock {A global collisionless PIC code in magnetic coordinates}.
\newblock {\em Computer Physics Communications}, 177(5):409--425, 2007.

\bibitem{Goerler_2011}
T.~{Goerler}, X.~{Lapillonne}, S.~{Brunner}, {Dannert} T., F.~{Jenko},
  F.~{Merz}, and D.~{Told}.
\newblock {The global version of the gyrokinetic turbulence code GENE }.
\newblock {\em {Journal of Computational Physics}}, 230(18):7053 -- 7071, 2011.

\bibitem{Sugama_2000}
H.~{Sugama}.
\newblock {Gyrokinetic field theory}.
\newblock {\em Physics of Plasmas}, 7(2):466--480, 2000.

\bibitem{brizard_prl_2000}
A.~J. {Brizard}.
\newblock {New Variational Principle for the Vlasov-Maxwell Equations}.
\newblock {\em Physical Review Letters}, 84(25):5768, 2000.

\bibitem{TBS_2016}
N.~{Tronko}, A.~{Bottino}, and E.~{Sonnendr\"ucker}.
\newblock {Second order gyrokinetic theory for Particle-In-Cell codes}.
\newblock {\em {Physics of Plasmas}}, 23:082505, 2016.

\bibitem{Brizard_2013}
A.~J. {Brizard}.
\newblock {Beyond linear gyrocenter polarization in gyrokinetic theory}.
\newblock {\em Physics of Plasmas}, 20(9):092309, 2013.

\bibitem{Tronko_Brizard_2015}
N.~{Tronko} and A.~J. {Brizard}.
\newblock {Lagrangian and Hamiltonian constraint for guiding-center Hamiltonian
  theories}.
\newblock {\em Physics of Plasmas}, 22(11):112507, 2015.

\bibitem{Cary_Brizard}
J.~R. {Cary} and A.~J. {Brizard}.
\newblock {Hamiltonian theory of guiding-center motion}.
\newblock {\em Reviews of Modern Physics}, 81(2):693--738, 2009.

\bibitem{Arnold_1989}
V.~I. {Arnold}.
\newblock {\em {Mathematical Methods of Classical Mechanics}}.
\newblock Springer, 1989.

\bibitem{Scott_Smirnov}
B.~{Scott} and J.~{Smirnov}.
\newblock {Energetic consistency and momentum conservation in the gyrokinetic
  description of tokamak plasmas}.
\newblock {\em Physics of Plasmas}, 17(11):112302, 2010.

\bibitem{Kleiber_2011}
R.~{Kleiber}, R.~{Hatzky}, A.~{Koenies}, K.~{Kauffmann}, and P.~{Helander}.
\newblock {An improved control-variate scheme for Particle-In-Cell simulations
  with collisions.}
\newblock {\em Computer Physics Communication}, 182:1005--1012, 2011.

\bibitem{newcomb1961lagrangian}
W.A. Newcomb.
\newblock Lagrangian and hamiltonian methods in magnetohydrodynamics.
\newblock Technical report, California Univ., Livermore (USA). Lawrence
  Livermore Lab., 1961.

\bibitem{Hahm_1988}
T.~S. {Hahm}.
\newblock {Nonlinear gyrokinetic equations for tokamak microturbulence}.
\newblock {\em Physics of Fluids}, 31(9):2670, 1988.

\bibitem{Hahm_Lee_1988}
T.~S. {Hahm}, W.~W. {Lee}, and A.~{Brizard}.
\newblock {Nonlinear gyrokinetic theory of finite-beta plasmas}.
\newblock {\em {Physics of Fluids}}, 31:1940, 1988.

\bibitem{Lee_1983}
W.W. {Lee}.
\newblock {Gyrokinetic approach in particle simulations.}
\newblock {\em Physics of fluids}, 26:556--562, 1983.

\bibitem{Goerler_Tronko_2016}
T.~{Goerler}, N.~{Tronko}, W.~A. {Hornsby}, A.~{Bottino}, R.~{Kleiber},
  C.~{Norcini}, V.~{Grandgirard}, F.~Jenko, and E.~{Sonnendr\"ucker}.
\newblock {Intercode comparison of gyrokinetic global electromagnetic modes}.
\newblock {\em {Physics of Plasmas}}, 23:072503, 2016.

\bibitem{Bottino_Sonnendruecker}
A.~{Bottino} and E.~{Sonnendr\"ucker}.
\newblock {Monte Carlo Particle-In-Cell methods for the simulation of the
  Vlasov-Maxwell gyrokinetic equations}.
\newblock {\em Journal of Plasma Physics}, 81(5):435810501, 2015.

\bibitem{PhD_Dominsky_2016}
J.~{Dominski}.
\newblock {\em EPFL PhD thesis}, (submitted), 2016.

\bibitem{Dominski_2017}
J.~Dominski, S.~Brunner, G.~Merlo, T.M. Tran, and L.~Villard.
\newblock An arbitrary wavelength solver for global gyrokinetic simulations.
  application to the study of fine radial structures on microturbulence due to
  non-adiabatic passing electron dynamics.
\newblock {\em Physics of Plasmas}, accepted 2017.

\bibitem{Mishchenko_Koenies_2005}
A.~{Mishchenko}, A.~{Koenies}, and R.~{Hatzky}.
\newblock {Particle simulations with a generalized gyrokinetic solver}.
\newblock {\em {Physics of Plasmas}}, 12:062305, 2005.

\end{thebibliography}
\end{document}